\documentclass[pra,aps,twocolumn,notitlepage,superscriptaddress,showpacs,nofootinbib]{revtex4-2}

\usepackage{enumerate,appendix}
\usepackage{amsmath, amsthm, amssymb,commath,braket}
\usepackage{color,calc,graphicx}
\usepackage[usenames,dvipsnames,svgnames,table,cmyk,hyperref]{xcolor}
\usepackage[colorlinks]{hyperref}
\hypersetup{
	colorlinks = true,
	urlcolor = {blue},
	citecolor = {magenta},
	linkcolor= {blue}
}

\usepackage{graphicx}
\usepackage{amsmath}
\usepackage{latexsym}
\usepackage{bbm}
\usepackage{array}
\usepackage{makecell}
\usepackage[charter,cal=cmcal,sfscaled=false]{mathdesign}
\usepackage{subfigure, epsfig}
\usepackage{graphicx}
\usepackage{amsmath}
\usepackage{color}

\newcommand{\cdummy}{\cdot}
\newcommand{\comma}{{,}}
\newcommand{\infixor}{\text{ or }}
\newcommand{\nobracket}{}
\newcommand{\nosymbol}{}

\newcommand{\tmop}[1]{\ensuremath{\operatorname{#1}}}

\def\bra#1{\langle#1|} \def\ket#1{|#1\rangle}

\def\proj#1{\ket{#1}\!\bra{#1}}

\newcommand{\be}{\begin{equation}}
	\newcommand{\ee}{\end{equation}}

\begin{document}
	
	\title{Proposals for ruling out the real quantum theories in an entanglement-swapping quantum network with causally independent sources}
	
	\author{Jian Yao}
	\affiliation{Shenzhen Institute for Quantum Science and Engineering and Department of Physics, Southern University of Science and Technology, Shenzhen, 518055, China}
	\affiliation{Guangdong Provincial Key Laboratory of Quantum Science and Engineering, Southern University of Science and Technology, Shenzhen, 518055, China}
	
	\author{Hu Chen}
	\affiliation{Shenzhen Institute for Quantum Science and Engineering and Department of Physics, Southern University of Science and Technology, Shenzhen, 518055, China}
	\affiliation{Guangdong Provincial Key Laboratory of Quantum Science and Engineering, Southern University of Science and Technology, Shenzhen, 518055, China}
	
	\author{Ya-Li Mao}
	\affiliation{Shenzhen Institute for Quantum Science and Engineering and Department of Physics, Southern University of Science and Technology, Shenzhen, 518055, China}
	\affiliation{Guangdong Provincial Key Laboratory of Quantum Science and Engineering, Southern University of Science and Technology, Shenzhen, 518055, China}
	
	\author{Zheng-Da Li}
	\email{lizd@sustech.edu.cn}
	\affiliation{Shenzhen Institute for Quantum Science and Engineering and Department of Physics, Southern University of Science and Technology, Shenzhen, 518055, China}
	\affiliation{Guangdong Provincial Key Laboratory of Quantum Science and Engineering, Southern University of Science and Technology, Shenzhen, 518055, China}
	
	\author{Jingyun Fan}
	\email{fanjy@sustech.edu.cn}
	\affiliation{Shenzhen Institute for Quantum Science and Engineering and Department of Physics, Southern University of Science and Technology, Shenzhen, 518055, China}
	\affiliation{Guangdong Provincial Key Laboratory of Quantum Science and Engineering, Southern University of Science and Technology, Shenzhen, 518055, China}
	\affiliation{Center for Advanced Light Source, Southern University of Science and Technology, Shenzhen, 518055, China}
	
	\begin{abstract}
		The question of whether complex numbers play a fundamental role in quantum theory has been debated since the inception of quantum mechanics. Recently, a feasible proposal to differentiate between real and complex quantum theories  based on the technique of testing Bell nonlocalities has emerged [Nature 600, 625–629 (2021)]. Based on this method, the real quantum theory has been falsified experimentally in both photonic and superconducting quantum systems [Phys. Rev. Lett. 128, 040402 (2022), Phys. Rev. Lett. 128, 040403 (2022)].
		The quantum networks with multiple independent sources which are not causally connected have gained significant interest as they offer new perspective on studying the nonlocalities. The independence of these sources imposes additional constraints on observable covariances and leads to new bounds for classical and quantum correlations. In this study, we examine the discrimination between the real and complex quantum theories with an entanglement swapping scenario under a stronger assumption that the two sources are causally independent, which wasn't made in previous works. Using a revised Navasc\'ues-Pironio-Ac\'in method and Bayesian optimization, we find a proposal with optimal coefficients of the correlation function which could give a larger discrimination between the real and quantum theories comparing with the existing proposals. This work opens up avenues for further exploration of the discrimination between real and complex quantum theories within intricate quantum networks featuring causally independent parties.
	\end{abstract}
	
	\maketitle
	
	\section{introduction}\label{sec1}
	
	Quantum mechanics, which established nearly 100 years ago, has achieved numerous significant accomplishments that have had profound impacts on science and technology \cite{dirac_principles_1981,von_neumann_mathematical_1955}. Especially in the past two decades, quantum information science, based on the principles of quantum mechanics and information science, has activated a series of advanced technologies such as quantum computing \cite{ladd2010quantum,arute2019quantum,zhong2020quantum,wu2021strong,zhu2022quantum,madsen2022quantum}, quantum communication \cite{gisin2007quantum,xu2020secure}, and quantum metrology \cite{giovannetti2011advances,polino2020photonic}. However, the fundamental role of complex numbers in quantum mechanics has long puzzled its founders and subsequent researchers \cite{realeinstein_letters_2011}. Despite this, a group of dedicated scientists has persistently pursued a quantum theory that relies on only real numbers in its mathematical formulation, running parallel to the development of the standard quantum theory \cite{realStueckelberg1960,realwootter_1981,realZurek1990-ZURCEA,Gisin_09_prl_simulate,realHardy_12,realAleksandrova_13,realMoretti_17,realDrechsel_19}. With a fixed Hilbert space dimension, the real and complex quantum theories can be discriminated by a single-site experiment using local tomography \cite{realGuo_Guang-Can_21PhysRevLett.126.090401}. However, without bounding the dimension, one should consider experiments involving several distant labs, such as the Bell non-locality test \cite{bellorigin}.
	
	Bell non-locality, was introduced by John Bell in 1964, with studying quantum correlations in a groundbreaking two-party model through the analysis of outcome statistics in experiments \cite{bellorigin}. Bell's model has made significant contributions to our understanding of quantum phenomena and has been instrumental in experimental tests that successfully ruled out local hidden variable theories \cite{Clauser_1972_PRL,Aspect_1982_PRL,belltest_Zeilinger_15PhysRevLett.115.250401,belltest_Hensen_2015,belltest_Shalm_15_PhysRevLett.115.250402,belltest_Weinfurter_17_PhysRevLett.119.010402,belltest_Pan_2018_PhysRevLett.121.080404}. In past decades, the studies of nonlocalities have been extended to the scenarios of quantum networks \cite{networkTavakoli_2022}, such as the nonbilocalities in an entanglement-swapping quantum network \cite{network_Gisin_2010_PhysRevLett.104.170401,network_Gisin_2012PhysRevA.85.032119}. These offers new opportunities and perspectives for studying quantum nonlocalities \cite{networkFritz_2012,network_Gisin_2012PhysRevA.85.032119,fritz2016beyond,network_Carvacho2017,network_Dylan_2017_doi:10.1126/sciadv.1602743,Miguel_2019_network,network_Sun2019,renou2019genuine,network_Poderini2020,network_Agresti_PRXQuantum.2.020346,network_Baumer2021,network_Tavakoli_2021_PhysRevLett.126.220401,network_Huang_2022_PhysRevLett.129.030502,pozas2022full,haakansson2022experimental,gu2023experimental,wang2023certification,Mao_2023_PRR,LigthartJMP2023,tavakoli2023semidefinite,brian2023maximalstar,Polino2023exnetwork}. In addition, quantum networks also play important roles in ruling out quantum theories based on only real numbers. It turns out that without bounding the dimension of Hilbert space, the real and complex quantum theories can not be discriminated without network-based scenario, even in a conventional Bell scenario with more than two separate parties \cite{Gisin_09_prl_simulate, DINPA_Moroder_2013, Pal_PRA_2008}. Recently, Renou et al., with the entanglement-swapping scenario \cite{Renounature}, give an interesting proposal to falsify quantum theory based on only real numbers. Their model involves three observers, namely Alice, Bob, and Charlie, and two independent entangled pairs without quantum correlations but allowing sharing a global hidden variable $\lambda$ as shown in Fig. \ref{fig1}(a). In their protocol, Bob performs a single joint measurement with four possible outcomes recorded as $b$, while Alice and Charlie perform three and six measurements with two outcomes with $x$, $z$ as the input and $a$, $c$ as the outcome, respectively. We refer to such a protocol as the (3,6)-scenario in this text. By defining a correlation function based on the outcome probability distributions from experimental data, Renou et al. \cite{Renounature} demonstrate that this function can assume different values under the constraints of real and complex quantum theories. These values can be calculated numerically using Semidefinite Programming (SDP) optimization techniques. This meticulously designed experiment has been conducted in both photonic and superconducting qubit systems and successfully falsified the quantum theory based on only real numbers with compelling experimental evidences \cite{expeFan,expePan,exPanstrict}. Following the work of Renou et al. \cite{Renounature}, Bednorz and Batle reduced the number of Charlie's measurement settings from six to four and three constructing the (3,3)- and (3,4)-scenario, and prove that a (2,2)-scenario does not exist \cite{pra}. In these models, a potential global hidden variable is taken into consideration, allowing for prior classical correlations among the parties. In quantum networks that incorporate multiple independent sources \cite{network_Gisin_2010_PhysRevLett.104.170401,networkFritz_2012,networkTavakoli_2022}, the causal structure becomes increasingly complex. The independence of these sources in various network structures introduces additional constraints on classical and quantum correlations. For a simplest entanglement-swapping quantum network with two casually independent sources involving two independent hidden variables $\lambda_1,~\lambda_2$ as shown in Fig. \ref{fig1}(b), there are constraints on the observers' measurement results coming from independence of two sources \cite{network_Gisin_2010_PhysRevLett.104.170401,network_Gisin_2012PhysRevA.85.032119}
	\begin{equation}
		\sum_b P (a, b, c|x, z) = P(a|x)P(c|z) = \sum_b P (a, b|x) \sum_b P (b, c|z) 
		\label{indeconst_0}
	\end{equation}
	where $x$, $z$ are the input of Alice and Charlie, while $a$, $b$, $c$ are the outcomes of Alice, Bob and Charlie. These new constraints bring a new maximum bound for numerically calculating a Bell type function under classical and standard quantum theories with a revised Navasc\'ues-Pironio-Ac\'in (NPA) method \cite{Miguel_2019_network} which introduce a scalar extension of the moment matrices. 
	
	In this work, we extend the study on the discrimination between real and complex quantum theories to the scenarios of quantum networks with multiple independent sources. Specifically, we focus on a simplest entanglement-swapping model involving two independent hidden variables. We introduce a revised NPA method, building upon the technologies used in previous works \cite{Renounature} and \cite{Miguel_2019_network} for numerical calculation of the maximum bounds of correlation function under the quantum theories that rely on only real numbers. Furthermore, we employ Bayesian optimization techniques to effectively search the optimal correlation function for discriminating the real and quantum theories, and successfully find a correlation function with a group of coefficients that exhibits superior robustness compared to existing proposals.

	\section{The entanglement swapping scenario}\label{sec2}
	
	\begin{figure}[h]
		\centering
		{\includegraphics[scale=0.37]{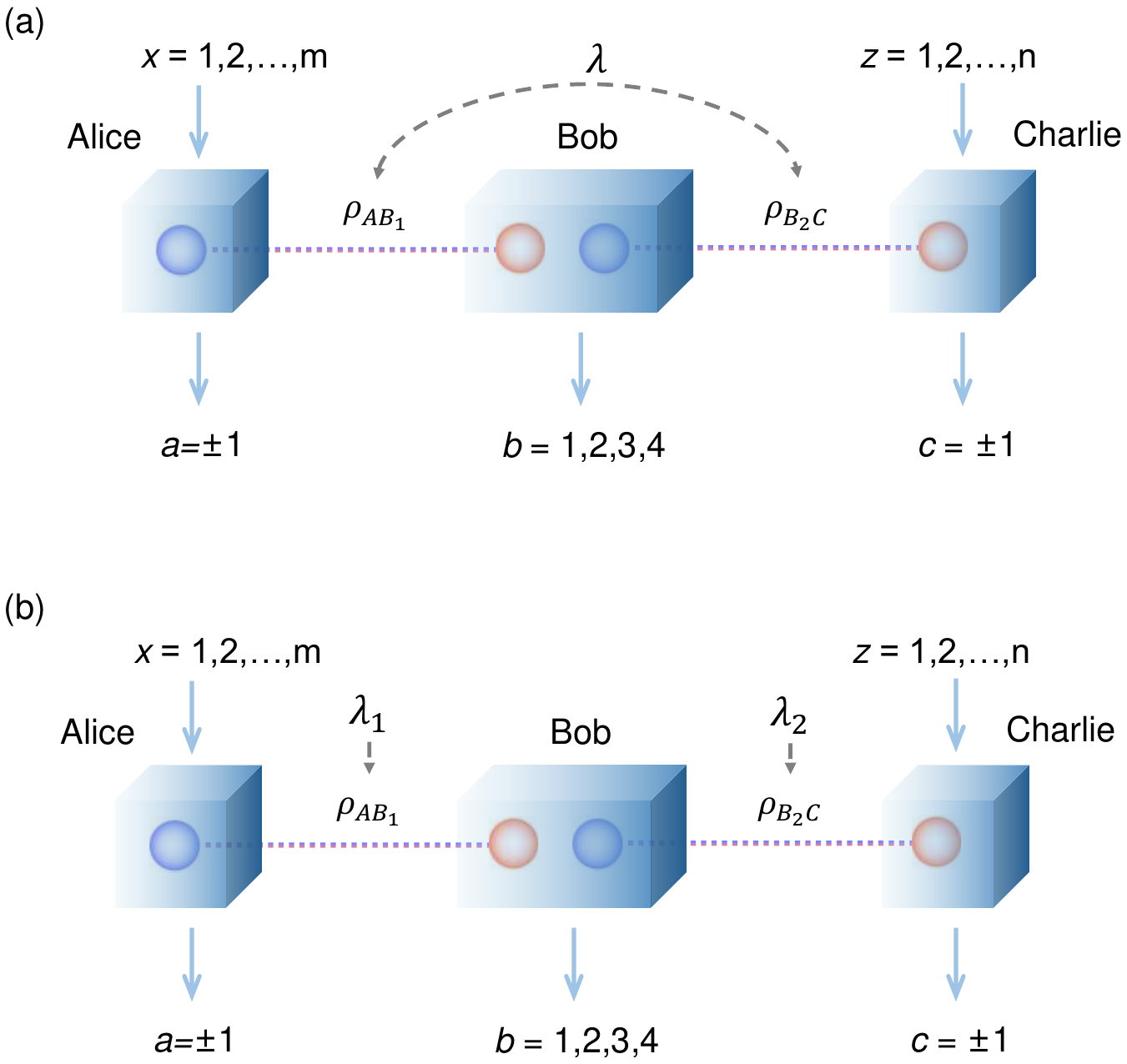}}
		\caption{Scenarios to discriminate complex and real quantum theories. $(a)$ The model used in \cite{Renounature} and \cite{pra}. $(b)$ The model used in this work, two independent hidden variables emphasize the causally independence of two sources. The $x$ and $z$ represent the input value of Alice and Charlie, while $a$, $b$ and $c$ represent the outcome of Alice, Bob and Charlie, respectively.}\label{fig1}
	\end{figure}
	
	The entanglement swapping scenario has been used to discriminate the real and complex quantum theories as discussed in Ref. \cite{Renounature} with the assumption that the two sources are allowed sharing a classical hidden variable. In this section we brief elaborate the new entanglement swapping model with causally independent sources. As shown in Fig. \ref{fig1}(b), such a scenario involves three observers, Alice, Bob, and Charlie, as well as two independent entanglement sources $\rho_{A B_1}$ and $\rho_{B_2 C}$. The entanglement source $\rho_{AB_1}$ is shared between Alice and Bob, while the entanglement $\rho_{B_2C}$ is shared between Bob and Charlie. Unlike the previous model used in \cite{Renounature} (Fig. \ref{fig1}(a)), these two sources are entirely causally independent, devoid of both quantum and classical correlations. They are characterized by two potential hidden variable with independent origin, denoted as $\lambda_1$ and $\lambda_2$.
	In this scenario, Bob conducts a single joint measurement on the two particles he received from the two entanglement sources obtaining four different outcomes recorded as $b =\{1, 2, 3, 4\}$, while Alice and Charlie randomly perform $m$ and $n$ dichotomic measurements on their received particles obtaining the outcomes $a=\{1,-1\}$ and $c=\{1,-1\}$, respectively. We refer to such a protocol as the ($m$,$n$)-scenario in this text. We use a group of conditional probabilities denote the experimental results. For example $P (a, b, c|x, z)$ represents the probability of outcome results $a, b, c$ when Alice, Bob and Charlie's measurement settings $A_x, B_b, C_z$, respectively, $x \in \{ 1, 2, ..., m \}$ and $z \in \{ 1, 2, ..., n \}$.
	
	In this work, we define an open correlation function with the outcome probability distribution and  an $m\times n$ free coefficient matrix $\mathcal{E}$ in the form of 
	\begin{equation}
		\begin{aligned}
			F &= \sum_{x, z, b} g (b, x) e_{x z} S_{x, z}^b, \\
			\tmop{where}~~ S^b_{x, z} &=
			\sum_{a, c \in \{ \pm 1 \}} a c P (a, b, c|x, z), \\
			g (b, x) &= \left\{
			\begin{array}{l}
				1, ~~~~\tmop{if}~ b = x \infixor b = 4\\
				- 1, ~~\tmop{otherwise}
			\end{array} \right.. 
		\end{aligned}
		\label{F}
	\end{equation}
	Here, $e_{x z}$ is the element of coefficient matrix $\mathcal{E}$. Similar to the Bell correlation function, the function $F$ may take different values depending on whether it is evaluated in the context of classical theory, real quantum theory, or complex quantum theory, and the maximum bound in each respective condition can be recorded as $F_c$, $F_r$ and $F_q$. If we observe a value of $F$ which exceeds $F_c$ and $F_r$ but not reaches $F_q$, we can falsify quantum theory based on only real numbers. The gap of $F_q$ and $F_r$ represent the discrimination between real and complex quantum theories. We define the value of $R = F_r / F_q$ represent the ratio of maximum bound of (\ref{F}) under real and complex quantum theory respectively. A smaller $R$ indicates a larger discrimination between real and complex quantum theories, and may reduce the requirements for the fidelities of the entanglement sources and measurements in practical experiments. For example, we discuss the case that three are white noises in two entanglement sources and the Bell state measurements Bob performed on the two particles he received. Then the entanglement source and Bell state measurement in complex quantum mechanics can be rewritten as 
	\begin{eqnarray}
		\rho_{A B_1}=\rho_{B_2 C} & = & v_E \proj{\Phi^+} + (1-v_E) I,\nonumber\\
		B_1 & = & v_I \proj{\Psi^+} + (1-v_I) I/4 ,\nonumber\\
		B_2 & = & v_I \proj{\Psi^-} + (1-v_I) I/4 ,\nonumber\\
		B_3 & = & v_I \proj{\Phi^+} + (1-v_I) I/4 ,\nonumber\\
		B_4 & = & v_I \proj{\Phi^-} + (1-v_I) I/4 ,
	\end{eqnarray}
	where $v_E$ and $v_I$ are the visibility of entanglement source and BSM respectively, $I$ is the Identity matrix. 
	The noised value $S'^b_{x,z}$ can be calculated as 
	\begin{equation}
		S'^b_{x,z} = \tmop{tr}{(\rho_{A B_1}\otimes\rho_{B_2 C})(A_x\otimes B_b\otimes C_z)} = v^2_E v_I S^b_{x,z}.
	\end{equation}
	Then the noised correlation function $F'_q=v^2_Ev_IF_q$. To experimentally falsify real quantum mechanics, it is necessary for us to satisfy the condition $v^2_Ev_I > F_r / F_q$. It is clear that a lower values of $R$ require lower visibility in experimental realization. To optimize the coefficient matrix $\mathcal{E}$ and attain a lower value of $R=F_r / F_q$, we employ Bayesian optimization, as outlined in Section \ref{sec4}.

	\section{Calculating the bound of real quantum theories in the new model involving two independent hidden variables}\label{sec3}

	In this section, we initially outline the process of formulating an SDP optimization problem for the computation of $F_r$ and $F_q$, primarily in accordance with the approach detailed in the reference \cite{Renounature}. Subsequently, for the new model used in this work, we demonstrate the incorporation of supplementary constraints from causal independence, as specified in Equation (\ref{indeconst_0}), into the SDP optimization problem.
		
	In the swapping scenario shown in Fig. \ref{fig1}(b), the probability distribution of measurement outputs can be represented by
	\begin{equation}
		P (a, b, c|x, z) = \tmop{tr} ( (\rho_{ABC}) (A_{a|x}
		\otimes B_b \otimes C_{c|z}) ), \label{prodis}
	\end{equation}
	where $\rho_{ABC}=\rho_{AB_1}\otimes\rho_{B_2C}$, and $A_{a|x}$ denotes the measurement operator we use when Alice chooses $x$ from possible settings and gets output $a$, notice that measurement operators and density operators here can live in either complex Hilbert space or real Hilbert space.
	
	To establish the constraints for Equation (\ref{prodis}) when computing the upper bound of Equation (\ref{F}), we employ methodologies previously employed in earlier works \cite{Renounature,expeFan,pra}, building upon Moroder et al.'s extension \cite{DINPA_Moroder_2013} of the NPA hierarchy \cite{NPA,navascues2008convergent,pironio2010convergent}. Our approach begins with the creation of two sets, $\mathcal{A}$ and $\mathcal{C}$, derived from the settings of Alice and Charlie. Specifically, $\mathcal{A}$ encompasses all monomials of ${ I, A_{1|1}, A_{1|2}, A_{1|3} }$ with degrees of $n_A$ or less (for the definition of monomial degree, refer to \cite{NPA}). Similarly, $\mathcal{C}$ is constructed following the same principle. Each monomial within $\mathcal{A}$ is linked to a ket denoted as $| \alpha \rangle$, with an associated property $\langle \alpha | \alpha' \rangle = \delta_{\alpha, \alpha'}$. Analogously, for monomials within $\mathcal{C}$, an orthonormal set ${ | \gamma \rangle }$ can also be associated. Then we define two completely positive map
	\begin{eqnarray}
		\Omega_A (\eta) & = & \sum_{\alpha, \alpha'} \tmop{tr} (A_{\alpha}^{\dagger}
		\eta A_{\alpha'}) | \alpha \rangle \langle \alpha' |, \nonumber\\
		\Omega_C (\eta) & = & \sum_{\gamma, \gamma'} \tmop{tr} (C_{\gamma}^{\dagger}
		\eta C_{\gamma'}) | \gamma \rangle \langle \gamma' |,  \label{twocp}
	\end{eqnarray}
	where $A_{\alpha}$ denotes the monomial that $| \alpha \rangle$ is  associated to, then we define a matrix
	\begin{equation}
		\Gamma^b = (\Omega_A \otimes \Omega_C) (\rho_{A C|b}), \label{gammab}
	\end{equation}
	where $\rho_{A C|b} = \tmop{tr}_B \{ (\rho_{A B_1} \otimes \rho_{B_2 C}) (I \otimes B_b \otimes I) \}$, is the reduced state of system $A$ and $C$ after Bob conducts measurement, obtaining output $b$.
	
	Since $\Omega_A$ and $\Omega_C$ are completely positive, $\Gamma^b$ is positive semidefinite. And it has other properties due to orthogonality of $\nobracket \{ | \alpha \rangle \}$ and $\nobracket \{ | \gamma \rangle \}$, that is, if $\alpha_2 \alpha_1^{\dagger} = \alpha_4 \alpha_3^{\dagger} = \alpha, \gamma_2 \gamma_1^{\dag} = \gamma_4 \gamma_3^{\dag} = \gamma$, we have
	\begin{equation}
		\langle \alpha_1 \gamma_1 | \Gamma^b | \alpha_2 \gamma_2 \rangle = \langle
		\alpha_3 \gamma_3 | \Gamma^b | \alpha_4 \gamma_4 \rangle
		= \tmop{tr} \{\rho_{A C|b} (A_{\alpha} \otimes C_{\gamma}) \}, 
		\label{orthorelation}
	\end{equation}
	which allows us to write $\Gamma^b$ in a more simple way
	\begin{equation}
		\Gamma^b = \sum_{\alpha \in \mathcal{A} \nosymbol \cdummy \mathcal{A},
			\gamma \in \mathcal{C} \cdummy \mathcal{C}} d^b_{\alpha, \gamma} M^{\alpha}
		\otimes N^{\gamma}, \label{simplegammab}
	\end{equation}
	where
	\begin{equation}
		M_{a, a'}^{\alpha} = \delta_{\alpha, a' a^{\dag}}, ~N^{\gamma}_{c, c'} = \delta_{\gamma, c' c^{\dag}},
	\end{equation}
	and the real coefficients $\{ d^b_{\alpha, \gamma} \}$ is the variable set of the optimization problem, it follows
	\begin{eqnarray}
		&&d^b_{A_{1| x}, C_{1| z}} = P (1, b, 1| x, z),~d^b_{A_{1| x}, I} = P (1,
		b|x), \nonumber\\
		&&d^b_{I, I} = P (b),~d^b_{\alpha, \gamma} = \tmop{tr} \{
		(A_{\alpha} \otimes B_b \otimes C_{\gamma}) (\rho_{A B C}) \}, \label{dandpro}
	\end{eqnarray}
	and the normalization constraint is $\sum_b d^b_{I, I} = 1$.
	
	Now we consider $\Gamma = \sum_b \Gamma^b$, by the independence of $\rho_{A B_1}$
	and $\rho_{B_2 C}$, we have
	\begin{equation}
		\Gamma = \Omega_A (\rho_A) \otimes \Omega_C (\rho_C),
	\end{equation}
	separability leads to different constraints in complex and real quantum theories \cite{PPT, realsp}
	\begin{equation}
		\left\{ \begin{array}{l}
			\Gamma^{T_A} \geq 0, ~\tmop{for} \tmop{complex} \tmop{quantum}
			\tmop{theory}\\
			\Gamma^{T_A} = \Gamma,~ \tmop{for} \tmop{real} \tmop{quantum} \tmop{theory}
		\end{array}, \right. \label{sepconst}
	\end{equation}
	the later constraint in Eq. (\ref{sepconst}) is stronger, leading to a lower upper bound of $F$.
	
	As we mentioned in sec. \ref{sec1}, in the new model involving $\lambda_1,~\lambda_2$ shown in Fig. \ref{fig1}(b), the independence of $\rho_{A B_1}$ and $\rho_{B_2 C}$ brings new constraints on probability distribution (notice that this formula is the same as Eq. (\ref{indeconst_0})):
	\begin{equation}
		\sum_b P (a, b|x) \sum_b P (b, c|z) = \sum_b P (a, b, c|x, z), \label{indeconst}
	\end{equation}
	however, due to their nonlinear nature, direct inclusion of the constraints from Eq. (\ref{indeconst}) into an SDP optimization problem is not feasible. To address this, we expand upon the technique presented in \cite{Miguel_2019_network}, adapting it to a bipartite form. This adaptation enhances compatibility with the aforementioned numerical approach, rendering it more suitable for implementation.
	
	We modify the set $\mathcal{C}$ constructed from Charlie's settings as (take (3,3) for instance)
	\begin{equation}
		\mathcal{C}= \{ I, C_{1|1}, C_{1|2}, C_{1|3}, \ldots, C_{1|2} C_{1|3}, c_1
		I, c_2 I, c_3 I \},
	\end{equation}
	where $c_i = P (c = 1| z = i) $, $\mathcal{A}$ does not change.
	
	$\{ d^b_{\alpha, \gamma} \}$ are constructed by the same process as Eq. (\ref{twocp}), Eq. (\ref{gammab}), Eq. (\ref{orthorelation}), and Eq. (\ref{simplegammab}), with more constraints coming from Eq. (\ref{indeconst})
	\begin{eqnarray}
		\sum_b d^b_{\alpha, c_i I} & = & \sum_b d^b_{\alpha, C_{1| i}}, ~\tmop{for}
		\forall \alpha,~\forall i \in \{ 1, 2, 3 \}, \nonumber\\
		\sum_b d^b_{I, c_i C_{1| j}} & = & \sum_b d^b_{I, c_i c_j I}, ~\tmop{for}
		\forall i,~j \in \{ 1, 2, 3 \},  \label{causalconst}
	\end{eqnarray}
	which can be derived from Eq. (\ref{dandpro}) and Eq. (\ref{indeconst}), specifically,
	\begin{eqnarray}
		\sum_b d^b_{\alpha, c_i I} & = & \sum_b \tmop{tr} \{ (\alpha \otimes c_i I)
		\rho_{A C|b} \} \nonumber\\
		& = & c_i \sum_b \tmop{tr} \{ (\alpha \otimes I) \rho_{A C|b} \}
		\nonumber\\
		& = & c_i \sum_b d^b_{\alpha, I} = \sum_b d^b_{\alpha, C_{1| i}}, 
	\end{eqnarray}
	the second formula in Eq. (\ref{causalconst}) can be derived in the same way.
	
	The new constraints Eq. (\ref{causalconst}) should also be satisfied under complex quantum theory. Hence, to calculate the \ maximum bound of $F$ (Eq. (\ref{F})) under real and complex quantum theory, we can construct and solve an SDP optimization problem,
	\begin{eqnarray}
		\max &  & F \nonumber\\
		s.t. &  & \Gamma^b = \sum_{\alpha \in \mathcal{A} \cdummy \mathcal{A} \comma
			\gamma \in \mathcal{C} \cdummy \mathcal{C}} d^b_{\alpha, \gamma} M^{\alpha}
		\otimes N^{\gamma} \geq 0, \nonumber\\
		&  & \sum_b P (b) = 1, \nonumber\\
		&  & \sum_b d^b_{\alpha, c_i I} = \sum_b d^b_{\alpha, C_{1| i}}, ~\tmop{for}
		\forall \alpha,~\forall i \in \{ 1, 2, 3 \}, \nonumber\\
		&  & \sum_b d^b_{I, c_i C_{1| j}} = \sum_b d^b_{I, c_i c_j I}, ~\tmop{for}
		\forall i,~j \in \{ 1, 2, 3 \}, \nonumber\\
		&  & \left(\sum_b \Gamma^b = \left( \sum_b \Gamma^b \right)^{T_A}  (\tmop{if}
		\tmop{real})\right), \label{SDP} 
	\end{eqnarray}
	the corresponding relationship between $\{ P (a, b, c|x, z) \}$ and $\{ d^b_{\alpha, \gamma} \}$ is illustrated by Eq. (\ref{dandpro}).
	 
	\section{Searching optimal correlation function with Bayesian optimization}\label{sec4}
	
	Bayesian optimization is a powerful technique for hyperparameter tuning, which involves finding the optimal values of a set of parameters for a given objective function. One popular implementation of Bayesian optimization is sequential model-based optimization (SMBO), the detailed principles of which are discussed in \cite{NIPS}. Here we briefly introduce SMBO and illustrate how to use it in the discrimination of complex and real theories.
	
	SMBO is an iterative algorithm that generates a model of the objective function in each iteration to identify the next set of parameters to test, ultimately finding the optimal set of parameters. To construct the objective function model, SMBO utilizes the parameter tuning history $H=(x_{1:i},f(x_{1:i}))$, here $x_{1:i}$ and $f(x_{1:i})$ denote parameters and corresponding function values obtained in $i$ times iterations. In the $(i+1)^{\tmop{th}}$ iteration, an acquisition function is employed to determine the next parameter set value $x_{i+1}$. This acquisition function balances exploitation, which involves selecting values close to the current most optimal parameter set, and exploration, which employs randomness to prevent falling into local optimal solutions. SMBO is terminated when the predefined number of iterations is reached. This algorithm is particularly beneficial when the objective function is a black-box function that is challenging to evaluate and does not have well-defined derivatives.
	
	In the discrimination between complex and real quantum theories, the parameter set to optimize is $\{e_{x z}\}$ in Eq. (\ref{F}), and the objective function is $R (\mathcal{E}) = F_r / F_q$, where $F_r$ and $F_q$ are maximum bound of Eq. (\ref{F}) under real quantum theory and complex quantum theory, respectively. Both $F_r$ and $F_q$ can be calculated by the SDP optimization problem constructed in sec. \ref{sec3} through MATLAB packages mosek \cite{mosek} and yalmip \cite{yalmip}, sometimes $F_q$ can be calculated analytically under certain assumptions. We use MATLAB Bayesian optimization package \cite{matlab} to realize SMBO on the objective function $R(\mathcal{E})$.
	
	In (3,3) scenario, the optimal values for $\{ e_{x z} \}$ we get are (permutation has been done to make results symmetric)
	\begin{equation}
		\mathcal{E}=\left(\begin{array}{ccc}
			0.31993 & 0.5 & - 0.5\\
			0.5 & 0.31933 & 0.5\\
			- 0.5 & 0.5 & 0.31933
		\end{array}\right),
	\end{equation}
	with $F_r = 2.1134, F_q = 2.3283, R (\mathcal{E}) = 0.9077$, The specific states and measurements to achieve $F_q$ are \cite{pra}
	\[ \{ A_x \} = \{ \sigma_i \}, \]
	\[ \{ C_z \} = \left\{ \sum_i c_z^i \sigma_i \right\},~c_z^i = - \frac{e_{i
			z}}{\sqrt{e_{1 z}^2 + e_{2 z}^2 + e_{3 z}^2}}, \]
	\[ B_1 = (I\otimes I - \sigma_X \otimes \sigma_X - \sigma_Y\otimes \sigma_Y - \sigma_Z\otimes \sigma_Z) /4; \]
	\[ B_2 = (I\otimes I  - \sigma_X \otimes \sigma_X + \sigma_Y\otimes \sigma_Y + \sigma_Z\otimes \sigma_Z) /4; \]
	\[ B_3 = (I\otimes I  + \sigma_X \otimes \sigma_X - \sigma_Y\otimes \sigma_Y + \sigma_Z\otimes \sigma_Z) /4; \]
	\[ B_4 = (I\otimes I  + \sigma_X \otimes \sigma_X + \sigma_Y\otimes \sigma_Y - \sigma_Z\otimes \sigma_Z) /4; \]
	\[ \rho_{A B_1} = \rho_{B_2 C} = | \Phi^+ \rangle \langle \Phi^+ |,~ | \Phi^+
	\rangle = \frac{1}{\sqrt{2}} (|00 \rangle + | \nobracket 11 \rangle), \]
	where $\{ \sigma_i \}$ are Pauli operators.
	
	Similarly, in (3,4) scenario, the optimal set of parameters we obtained are 
	\begin{equation}
		\mathcal{E}=\left(\begin{array}{cccc}
			-0.19883 & 0.1996 & 0.20026 & 0.19944\\
			0.20094 & -0.19971 & 0.20083 & 0.1987\\
			0.2006 & 0.19961 & -0.2 & 0.19971
		\end{array}\right),
	\end{equation}
	we can infer that absolute values of all coefficients are the same, hence the optimal set is (with every coefficient timing 5)
	\begin{equation}
		\mathcal{E}=\left(\begin{array}{cccc}
			-1 & 1 & 1 & 1\\
			1 & -1 & 1 & 1\\
			1 & 1 & -1 & 1
		\end{array}\right),
	\end{equation}
	with $F_q = 6.9282, F_r = 6.4722, R (\mathcal{E}) = 0.8847$, which is lower than $R (\mathcal{E})$ of a (3,6) scenario in \cite{Renounature}, to achieve $F_q$, $\{ A_a \}$ and $\{ B_b \}$ used are the same as what (3,3) scenario uses, $(C_z)$ are
	\begin{equation}
		\left(\begin{array}{c}
			C_1\\
			C_2\\
			C_3\\
			C_4
		\end{array}\right) = \frac{1}{\sqrt{3}} \left(\begin{array}{ccc}
			1 & - 1 & - 1\\
			- 1 & 1 & - 1\\
			- 1 & - 1 & 1\\
			-1 & -1 & -1
		\end{array}\right) \left(\begin{array}{c}
			\sigma_X\\
			\sigma_Y\\
			\sigma_Z
		\end{array}\right) .
	\end{equation}
In the new swapping-entanglement model involving $\lambda_1,~\lambda_2$, new constraints (\ref{causalconst}) are included, so we obtain tighter upper bounds for $F_r$, Since $F_q$ at least has the same forms as previous work \cite{Renounature, pra}, we achieves lower values of $R(\mathcal{E})$. We compare the results presented in this work with those from previous research, as illustrated in Table \ref{table1}. The findings indicate that we have achieved lower values of $R(\mathcal{E})$, implying reduced visibility requirements for experimental implementation and hence exhibiting superior robustness.
	
\begin{table}[h]
	
	\begin{tabular}{|c|c|c|c|}
		\hline
		Scenario & $R (\mathcal{E})$ & Causal constraint\\
		\hline
		(3,3) \cite{pra} & 0.9381 & No\\
		\hline
		(3,3) [This work]& 0.9077 & Yes\\
		\hline
		(3,4) \cite{pra} & 0.9341 & No\\
		\hline
		(3,4) [This work]& 0.8847 & Yes\\
		\hline
		(3,6) \cite{Renounature} & 0.9028 & No\\
		\hline
	\end{tabular}
	\caption{Comparison between this work and previous works \cite{Renounature,expeFan,pra}\label{table1}}
\end{table}

\section{summary and discussions}
In summary, our work offers proposals to discriminate real and complex quantum theories in a new entanglement-swapping model involving two independent hidden variables, which emphasize the causally independent nature of sources. To address new constraints of the casual independence on the experimental outcome probability distribution, we have developed a numerical method based on the NPA technologies, enabling discrimination between real and complex quantum theories in quantum networks with causally independent parties. In this work, we further employ Bayesian optimization to search for the optimal coefficient matrix of the correlation function. As a result, we obtain a more experimentally feasible scenario that allows discriminating real and quantum theories with lower visibility of the sources and measurements, comparing to existing proposals. 

Finally, we discuss the reasonableness of the independence assumption in our model. Renou et al's and other previous models \cite{Renounature,expeFan,pra} consider a global hidden variable shared between the two sources in the entanglement-swapping scenario. This is because that the two entangled sources may be produced in the same factory or being operated using the same power socket. The causal network with an additional source should be considered more general than the one without such source. However, a very similar assumption is actually needed in Renou et al's and other previous models, that is the independence of randomness sources for choosing measurements, which also corresponds the “free choice” or “measurement independence” assumption in standard Bell tests \cite{network_Gisin_2010_PhysRevLett.104.170401,network_Gisin_2012PhysRevA.85.032119}. The stronger assumption that the entanglement source are absolutely independent, brings new constraints on the probability distribution of the input-output experiment as the form of Eq. \ref{indeconst_0}. These constraints are not only added on the experimental results predicted by complex quantum theory, but also added on the cases predicted by real quantum theories and classical theory. These constraints may change the effective probability distributions for all three cases, and help us find a more experimentally feasible scenario with lower demand of visibility of the sources and measurements. These considerations should be extended to exploring discrimination between real and complex quantum theories in more complicated quantum networks with causally independent parties which will introduce more additional constraints \cite{networkTavakoli_2022}. We believe further advancements for the discrimination of real and complex quantum theories will be achieved, ultimately making experiments more accessible in the near future.
		
\section*{acknowledgments}

We thank Adam Bednorz, Sixia Yu for helpful discussions. This work is supported by the Shenzhen Science and Technology Program Grant No.RCYX20210706092043065, Shenzhen Fundamental Research Program Grant No. JCYJ20220530113404009, the National Natural Science Foundation of China Grants No.12005090, No.62375117 and No.92365116, the Key-Area Research and Development Program of Guangdong Province Grant No.2020B0303010001 and No.2019ZT08X324, Guangdong Provincial Key Laboratory Grant No.2019B121203002 and SIQSE202104.

\bibliography{RQT}
			
\end{document}